\documentclass[12pt,preprint]{emulateapj}
\begin{document}

\newcommand{\etal}{{\it et al.}}

\slugcomment{Accepted for publication in ApJ}

\title{The first {\it Swift} X-ray Flash: The faint afterglow of XRF~050215B}

\author{
A.J. Levan\altaffilmark{1,2},
J. P. Osborne\altaffilmark{1},
N.R. Tanvir \altaffilmark{1,2},
K.L. Page\altaffilmark{1},
E. Rol\altaffilmark{1},
B. Zhang\altaffilmark{3},
M.R. Goad\altaffilmark{1},
P.T. O'Brien\altaffilmark{1},\\
R.S. Priddey\altaffilmark{2}
D. Bersier\altaffilmark{4}, 
D.N. Burrows\altaffilmark{5}, 
R. Chapman\altaffilmark{2}, 
A.S. Fruchter\altaffilmark{6}, 
P. Giommi\altaffilmark{7}, 
N. Gehrels\altaffilmark{8},\\
M.A. Hughes\altaffilmark{2},
S. Pak\altaffilmark{2},
C. Simpson\altaffilmark{9},
G. Tagliaferri\altaffilmark{10},
E. Vardoulaki\altaffilmark{11}
}

\altaffiltext{1}{Department of Physics and Astronomy, University of Leicester, University Road,
Leicester, LE1 7RH, UK}
\altaffiltext{2}{Department of Physics, Astronomy and Mathematics , University of Hertfordshire, College Lane, Hatfield, Herts, AL9 10AB, UK}
\altaffiltext{3}{Department of Physics, University of Nevada Las Vegas, Las Vegas, NV 89154}
\altaffiltext{4}{Astrophysics Research Institute, Liverpool John Moores University, Twelve Quays House, Birkenhead CH41 1LD}
\altaffiltext{5}{Department of Astronomy and Astrophysics, Penn State University, University Park, PA 16802, USA }
\altaffiltext{6}{Space Telescope Science Institute, 3700 San Martin Drive, Baltimore, MD21218, USA}
\altaffiltext{7}{ASI Science Data Center, Via Galileo Galilei, I-00044 Frascati, Italy }
\altaffiltext{8}{NASA/Goddard Space Flight Center, Greenbelt, MD 20771, USA}
\altaffiltext{9}{ Astrophysics Research Institute, Liverpool John Moores University, Twelve
Quays House, Egerton Wharf, Birkenhead CH41 1LD, UK}
\altaffiltext{10}{INAF-Osservatorio Astronomico de Brera, Via E. Bianchi 46, I-23807,
Merate (LC), Italy}
\altaffiltext{11}{Department of Physics, University of Oxford, Denys Wilkinson Building,
Keble Road, Oxford OX1 3RH}

\begin{abstract}
We present the discovery of XRF~050215B and its afterglow. The burst was
detected by the {\it Swift} BAT during the check-out phase and observations
with the X-ray Telescope began approximately 30 minutes after the burst.
These observations found a faint, 
slowly fading X-ray afterglow near the centre of the error box as reported by the BAT. 
Infrared data, obtained at UKIRT after 10 hours also revealed a 
very faint K-band afterglow. The afterglow appears unusual since it is very faint, especially
in the infrared, with K$>$ 20 only 9 hours post-burst. 
The X-ray and infrared lightcurves exhibit a slow, monotonic decay 
with $\alpha \sim 0.8$ and no evidence 
for a steepening associated with the jet break to 10 days  post burst. 
We discuss possible explanations for the faintness and slow decay
in the context of present models for the production of X-ray Flashes.
\end{abstract}


\section{Introduction}
X-ray Flashes (XRFs) appear to be a subclass of Gamma-Ray Bursts (GRBs). 
They have similar durations to the long-soft GRBs, but they have a low 
gamma-ray flux, a high ratio of X-ray to gamma-ray flux, and a spectral peak 
at a much lower energy (Kippen \etal\ 2003). 
First identified by the Wide Field Cameras on BeppoSAX 
(Heise \etal\ 2001) they have been located with 
increasing frequency by {\it HETE-2}. Followup observations have in some cases
successfully found afterglow emission, most commonly at X-ray wavelengths, and
occasionally in the optical and nIR. 
Accurate positions have indicated that, like GRBs, XRFs
are found in star-forming galaxies at cosmological distances (e.g. Bloom \etal\ 2003),
but, based on statistics of a few, at rather lower median redshift than GRBs.
Recently the very low-redshift XRF 060218, 145 Mpc distant, was shown
to be associated with SN 2006aj (Campana \etal\ 2006; Pian \etal\ 2006;  Modjaz \etal\ 2006), clearly
demonstrating that some XRFs, as for long duration GRBs originate in
the core collapse of a massive star in a type Ic supernova (Hjorth et al. 2003; Stanek et al. 2003). 
Photometric evidence for associated supernovae has also been seen in a number
of XRFs previously (e.g. Fynbo \etal\ 2004; Soderberg \etal\ 2004;2005; Bersier \etal\ 2006), 
however the absence of SN signatures in some cases indicate that
the SN emission may be markedly fainter than the prototypical GRB supernova SN~1998bw
(Levan \etal\ 2005; Soderberg \etal\ 2005). 

The existence of XRFs shows that the range of spectral properties associated with 
GRBs is very large. The peak of the $\nu F_{\nu}$ spectrum ($E_p$) can be seen from 
$<$5 keV to $>1$ MeV. An important
question is, therefore: 
what physical processes are responsible for this range of 
energies?
Observations of the prompt and afterglow emission of XRFs can be used
to probe this question and place constraints on the energy generation and physical
structure within these highly energetic explosions. 

Various scenarios have been proposed to explain the lower peak energy of
XRFs compared to GRBs; these can be split into several subsets of model, of 
which the most broadly discussed are: (i) 
GRBs at very high redshifts would have
their spectral energy distributions (SEDs) shifted into the X-ray window (Heise \etal\ 2001);
(ii) GRBs whose Lorentz factor is modified due to the effects
of baryon loading within the jet. In external shock models high
baryon loading (the so-called dirty fireball) can create an XRF
 (Dermer \etal\ 1999;
Ramirez-Ruiz \& Lloyd-Roming 2002; Huang \etal\ 2002), 
while in contrast for internal shock
models very clean jets produce large X-ray fluxes (Zhang \& Meszaros 2002,
Barraud \etal\ 2005). (iii) GRBs with either a differing geometry or 
observer viewing angle can also naturally create XRFs, for example
due to broader than normal jets (Huaag \etal\ 2004; Zhang \etal\ 2004);
bursts where the line of sight is
"off-axis" with respect to the jet orientation or bursts whose jet is structured, 
either in a simple (two component) manner, or with more complex 
variations with viewing angle. (iv) Finally it is possible that XRFs
represent manifestations of differing physical processes to GRBs themselves,
perhaps originating from a hot photosphere (Meszaros \etal\ 2002).


The discovery of a correlation between $E_p$ and the square root of
the isotropic energy release ($E_{iso}$) (Amati \etal\ 2002) directly implies
that softer bursts have lower energies and can be well explained by
models where the XRF is the result of a viewing angle effect.  In other words,
XRFs are seen when a classical GRB is viewed off the primary collimation so
the highest energy emission is not seen (e.g. Yamazaki \etal\ 2002; Rhoads 2003; Dado
\etal\ 2004).  However, outliers to this relation can be found (e.g. Sazonov et al. 2004),
and it remains to 
be confirmed that it is purely the result of a viewing angle effect. 

The {\it Swift} satellite was launched in November 2004 and is now delivering
localizations for 
approximately 2 GRBs per week (see Gehrels \etal\ (2004) for a 
description of the {\it Swift} mission). The passband of the GRB detectors on 
{\it Swift} is smaller than that on {\it HETE-2}, with a low energy response down
to 15 keV compared with 5 keV for {\it HETE-2}. XRF~050215B was the first XRF
to be detected by {\it Swift} and to have moderately rapid X-ray followup (previous XRFs have
not been observed in X-rays for at least several hours, and more typically days).
 Here we report the results of optical/IR and
X-ray observations of XRF~050215 and the constraints they place on models for the
production of XRFs.

\section{Observations}

XRF~050215B was detected by the {\it Swift} Burst Alert Telescope at 
02:33:12 UT on 15th February 2005; the initial location was 
RA = 11:37:48, Dec = 40:48:18 with a 90\% error
radius of 4 arcminutes (trigger 106106, Barthelmy \etal\ 2005a). 
{\it Swift} slewed to the burst promptly, but was unable to observe until 
$\sim$ 29 minutes after the trigger because it was in the high radiation 
environment of the South Atlantic Anomaly. 
X-ray Telescope observations revealed a faint, slowly fading point 
source, identified as the X-ray afterglow of XRF~050215B (Goad \etal\ 2005).
The UV and Optical Telescope did not yield
a detection of any source at the location of this 
X-ray object (Roming \etal\ 2005a,b).   

The burst was also seen by {\it HETE-2}, and an analysis of this data revealed a best fitting
$E_p$ = 17.6 keV, with a 95\% upper bound of $E_p < 30.3$ keV (Nakagawa \etal\ 2005). 
The ratio of fluxes in the 2-30 and 30-400 keV bands was 1.65, implying significantly more
X-ray emission than $\gamma$-ray and indicating that 050215B was indeed an XRF.

Initial ground-based observations failed to yield an optical counterpart. 
The BOOTES-2 wide field camera was observing the field at the time of the 
burst, limiting the unfiltered magnitude of any flash to $>10$ 
(Jelinek \etal\ 2005). 
ROTSE-III observations started 100s after the burst, however these images 
failed to yield an optical counterpart to limiting magnitudes of 
R$\sim 17.8 - 18.8$ up to 50 minutes after the burst  (Yost \etal\ 2005). 
Radio upper limits of 93 and 156 $\rm \mu$Jy (3 $\sigma$) at 2.9 and 12 days 
after the burst were reported by Soderberg \& Frail (2005a,b). A skymap with
the location of the burst (and afterglow) is shown in Figure 1. 


\begin{figure*}
\includegraphics[scale=1.1,angle=270]{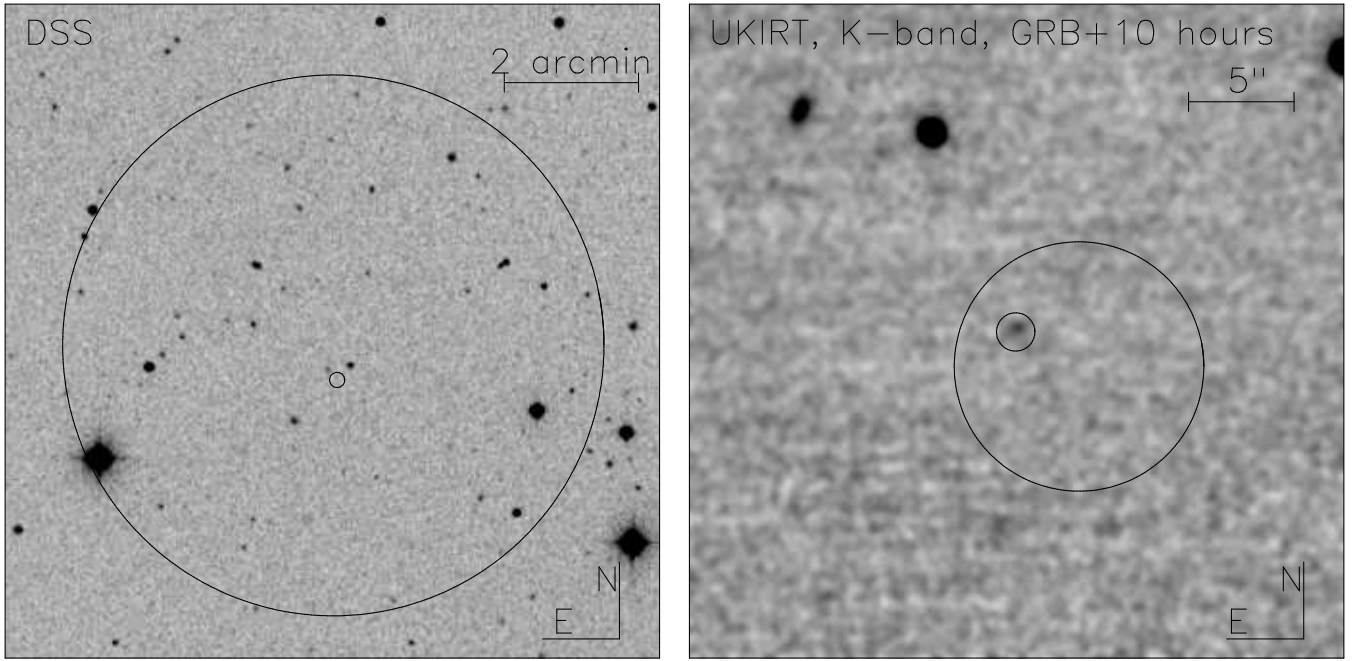}
\caption{The sky position of XRF~050215B. The left hand image shows a region of
the Digital Sky Survey (DSSII-red) with the BAT error box marked as the large circle and the
XRT position labeled within it. The right hand image shows our first UKIRT observation. 
The large circle in this image is the XRT location and 90 \% confidence region. The
location of the IR afterglow is also marked. }
\label{posn}
\end{figure*}

\subsection{{\it Swift} Observations}

The {\it Swift} satellite and its gamma-ray Burst Alert Telescope (BAT), 
X-ray Telescope (XRT) and UV/Optical Telescope (UVOT) 
are described by Gehrels \etal\ (2004), Barthelmy \etal\ (2004, 2005b), 
Burrows \etal\ (2004, 2005) and Roming \etal\ (2004, 2005c). 
The {\it Swift} data described here 
has been processed with version 2.0 of the {\it Swift} software\footnote{http://heasarc.gsfc.nasa.gov/docs/software/lheasoft/}, using  
the standard BAT and XRT pipelines. 

The BAT detected XRF~050215B below 150 keV, showing a single peaked 
lightcurve, (see Figure 2).
The rise to peak lasts 2-4 s, while the decline takes 7-10 s. 
The T$_{50}$ and T$_{90}$ and durations
are 3.4 and 7.8 seconds, respectively.
The BAT trigger time (02:33:43 UT) corresponds to the peak of the light curve. 
This time was mis-reported initially; all times from the burst stated in this 
paper use this correct trigger time.
A single powerlaw spectral fit to the entire burst 
between 15 and 150 keV results in a good fit 
($\chi^{2}/$dof $ = 1.0$), with
a photon power law index of 2.3$\pm$0.4. 
The total 15-150 keV fluence in
the T90 interval is $2.33 \pm 0.6 \times 10^{-7}$ erg cm$^{-2}$ (errors are quoted
as 90\%).
The extrapolated fluence in the 30-400 keV FREGATE region is 
2.0 $\times 10^{-7}$ erg cm$^{-2}$, in agreement with the results obtained 
by Nakagawa \etal (2005). While XRF~050215B had a fluence somewhat below the 
Swift average, it is unusual in having a very soft spectrum 
(corresponding to a 2-30 keV to 30-400 keV extrapolated fluence ratio of 
$S_X /S_{\gamma} = 2.0^{+0.2}_{-0.3}$) 
consistent with the X-ray flash characterisation of Nakagawa \etal (2005). 

Observations with the XRT
began at 03:03:11 UT (approximately 30 minutes after trigger), 
and continued for 13.3 days; 
they are detailed in table 1. The first exposure was made in 
Windowed Timing (WT) mode while all later observations were in 
Photon Counting (PC) mode; a total of 128 counts were obtained from 
XRF~050215B.
The location of the X-ray afterglow was found to be
RA = 11:37:47.66, Dec = 40:47:46.7 , 
with a 90\% error radius of 4.4 arcseconds (Moretti \etal\ 2006). 
The X-ray lightcurve of XRF~050215B is shown in Figure 3 (and
is combined with the BAT lightcurve). 
The WT data were extracted from a box 94$''$ across (40 pixels), with the background 
measured from the same size region offset from the source. For the PC mode 
data counts were extracted from a 47$''$ (20 pixel) radius region centred on the source, 
with the background taken from a source-free region 9 times larger in area. 
All points in Figure 3 correspond to a better than 3$\sigma$ detection of the 
source. The X-ray afterglow of XRF~050215B declines as an unbroken power law 
with a decay index of 0.82$\pm$0.08.
With the modest number of XRT counts only limited spectral information can 
be derived. We fit a power law model to the total PC mode afterglow spectrum 
(which covers 0.5-2.6 keV), finding a good 
fit for a photon index of 1.5$\pm$0.5 assuming the Galactic absorbing column 
of $2\times 10^{20}$ cm$^{-2}$. No excess absorption over that due to our
Galaxy is seen; at zero redshift $\Delta N_H < 3.8\times 10^{21}$ cm$^{-2}$
(90\%).
Examination of the ratio of counts in the 
0.3-1.0 keV to 1.0-10 keV bands as a function of time showed 
no variation. Unabsorbed 0.3-10 keV fluxes at 1 and 11 hours derived from 
Figure 3 and the spectral parameters above are respectively
28 and 4.9 $\times 10^{-13}$ ergs s$^{-1}$ cm$^{-2}$.

The {\it Swift} UVOT also observed the position of XRF~050215B at times 
coincident with the XRT observations using its full range of filters. 
No new source was detected, either initially or later with longer exposures. 
3 $\sigma$ upper limits include $U>20.2, B>20.1, V>19.3$ at 0.57, 0.60 , 0.48 
hours, and $V>21-22$ up to 13.3 days after the burst.

\begin{deluxetable*}{llllll }
\footnotesize
\tablecolumns{4}
\tablewidth{0pt}
\tablecaption{{\it Swift} XRT observations of XRF~050215B}
\tablehead{\colhead{Start (hours)} & \colhead{End (hours)} &
\colhead{Exp. Fraction} & \colhead{Source counts} &  \colhead{Background counts}
& \colhead{XRT mode}}
\startdata
0.498  &  0.656   & 1.0    & 45 & 16   & WT\\
1.603  &  1.699   & 1.0    & 11 & 0.22 & PC\\
3.219  &  3.416   & 1.0    & 17 & 0.56 & PC\\
4.819  &  5.133   & 1.0    & 16 & 0.67 &PC\\
6.419  &  6.783   & 1.0    & 14 & 0.89 & PC\\
8.036  &  8.399   & 1.0    & 13 & 0.56 & PC\\
47.048 &  66.455  & 0.081  & 16 & 4.6  & PC\\
69.528 &  93.433  & 0.14   & 19 & 12   & PC\\
95.250 & 320.572  & 0.0098 & 28 & 16   & PC\\
\hline
\enddata
\tablecomments{Swift XRT observations of XRF~050215B. A final observation at 13.3 
days was too short to be useful, and is not listed here. The start and end times are
given in hours since the burst (i.e. hours after 2005-02-15 02:33:12 UT)}
\label{tab:photdata}
\end{deluxetable*}

\subsection{Ground based observations}

IR images were obtained at UKIRT using the UFTI instrument 
(Roche \etal\ 2002) in the K98 filter (see Figure \ref{posn}). 
Dithered observations were flat-fielded, sky- and
dark-subtracted and combined using ORAC-DR (Cavanagh \etal\ 2002). 
IR and optical images were also taken at Gemini: NIRI data 
(Hodapp \etal 1995) 
which were also processed with ORAC-DR; GMOS optical data (Hook \etal 2004)
which were reduced via
the specific Gemini/GMOS tasks with IRAF \footnote{For information
on ORAC-DR see http://www.oracdr.org and for the Gemini IRAF tasks
see http://www.gemini.edu/sciops/data/dataIRAFIndex.html}. 
Additionally a 4 x 30 minutes spectrum of the potential host was taken with 
GMOS
with an R400 grating and the central wavelength varying between 6400
and 6550 Angstrom, from March 11.47 to 11.56 UT. Data reduction was
done in a standard fashion using IRAF. Only a very weak 
continuum was seen, there were no obvious emission lines  
in the combined spectrum between 6000 and 8500 Angstroms.
A complete log of observations is shown in
table \ref{phot}.

Our first image was obtained approximately 10 hours after the burst 
with a second epoch obtained the following night ($\sim$ 33 hours 
post-burst). Comparison of these images 
revealed a fading point source at RA=11:37:47.90, Dec = 40:47:45.6, 
approximately 33 arcseconds from the
centre of the BAT error box and well within the refined XRT position 
given in section 2.1.

To perform photometry of the afterglow we used an aperture equal to the FWHM of 
the images at each epoch. During our UKIRT observations photometric standards 
were also observed which allowed us to subsequently calibrate the field. The
photometric calibration of the R-band images was done using the photometry 
from the Sloan Digital Sky Survey, which covers the region of the burst in the same
filter as used for the GMOS observations.

Photometry of XRF~050215B is shown in table \ref{phot} and the K-band light curve 
is shown in Figure \ref{klc}. The afterglow was very faint, and therefore the signal to 
noise was typically low ($3-10 \sigma$ in most observations). However the afterglow
is clearly fading. A single power-law fit to the K-band data yields a best fit decay
slope (fitted as $F(t) \propto t^{-\alpha}$) of $\alpha = 0.47 \pm 0.08$, unusually slow for
GRB afterglow which fall typically with $\alpha = 1$ in the first hours to days after the burst, and
more rapidly after this. One possibility is that our measured fluxes do not contain pure afterglow
but some contribution from an underlying host galaxy. In this case the true decay slope may
be faster. We can estimate the maximum possible contribution from an underlying host by 
making the assumption that the host galaxy contributes a flux equal to the upper limit of
the final epoch observation (i.e. lies just below the detection threshold). Given
the R magnitude of the host (see below) and the typical R-K colours of GRB host galaxies 
(in the range 2.5 - 3, Le Floc'h \etal\ 2003) it is unlikely that the host 
is significantly fainter than this, and therefore this may provide an acceptable estimate
of the afterglow decay. The (putative host) subtracted photometry is shown in
red in Figure \ref{klc}. This photometry, while very uncertain, is
marginally consistent with a single power-law decay of index $\alpha = 0.82 \pm 0.08$;
thus, within the uncertainty (due to the unknown host magnitude),
the optical and X-ray slopes are consistent with being identical. 


PSF matched image subtractions of the optical observations taken 11, 21 and 115 days
after the burst reveal no evidence for any variation, demonstrating that the optical 
light from XRF~050215B was dominated by the host galaxy from 11 days onwards, 
and did not contain a significant contribution from either afterglow or associated supernova
(the limit on any transient emission at the location of the afterglow is R$\sim$25.8). 
The magnitude of this host galaxy is R=24.7, comparable to the median of GRB
host galaxies seen to date. Our final K-band point can also be used to place a limit
on the host colour. Assuming it has K$>$22.25 this implies that R-K$<$ 2.5. This is again
comparable to the R-K colours of GRB hosts which have a mean R-K=3 (Chary \etal\ 2002;
Le Floc'h \etal\ 2003).

Under the assumption that a supernova was associated with XRF 050215B, we can
use the observed limits on the supernova emission to crudely estimate a
lower bound to the redshift (also assuming that there is no excess extinction
along the line of sight to XRF 050215B). The limiting magnitude of
any supernova of R=25.8 is comparable to the peak magnitude of a
supernova such as SN~1998bw at $z=1$. However
more typical type Ic supernova such as those putatively associated
with XRFs (e.g. Fynbo \etal\ 2004; Bersier \etal\ 2005) span a range
of magnitudes from similar to SN~1998bw to a factor of 10 fainter,
these fainter supernovae would only be visible to moderate $\sim 0.5$ redshift.

\begin{figure}
\includegraphics[scale=0.3,angle=270]{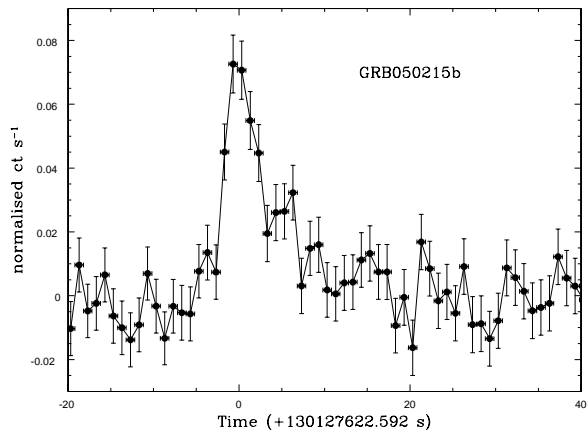}
\caption{The BAT 15-350 keV light curve of the XRF~050215B.}
\label{bat}
\end{figure}

\begin{figure}
\includegraphics[scale=0.35,angle=270]{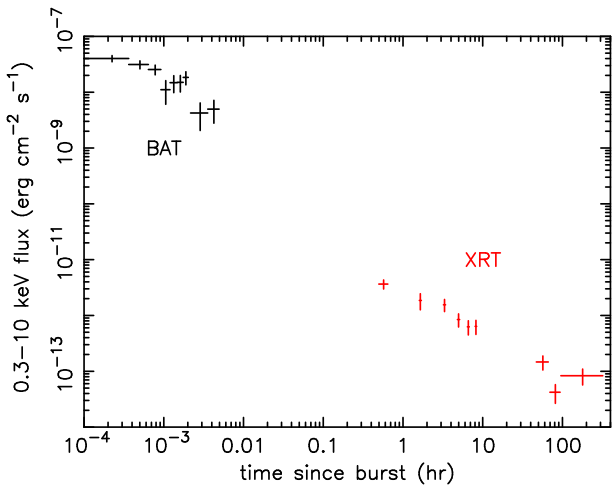}
\caption{The joint BAT and XRT lightcurve of XRF 050215B. 
The XRT count-rates were converted into unabsorbed fluxes using the best
fit model. The BAT data were extrapolated into the 0.3-10 keV band
assuming a photon index which was the mean of the best-fit XRT and BAT
spectral indices. The XRT data are best fit with an unbroken
power-law with a decay index $\alpha = 0.82 \pm 0.08$.
The extrapolation of the XRT observations to early
times falls slightly below the prompt emission measured by the BAT
(and the fitting of a single power-law to both BAT and XRT data does not result in a
statistically acceptable fit, although given the calibration
uncertainties it may be consistent with a single
decline from $10^{-3}$ hours onward). However, an extrapolation of the X-ray afterglow
below the prompt emission could easily be explained by a steep decay
of the X-ray flux at early times
 which is
commonly seen in {\it Swift} bursts (e.g. Tagliaferri \etal\ 2005; Nousek \etal\ 2006;
O'Brien \etal\ 2006).}
\label{xrtlc}
\end{figure}


\begin{deluxetable*}{lllllll}
\footnotesize
\tablecolumns{4}
\tablewidth{0pt}
\tablecaption{Ground based observations of XRF~050215B}
\tablehead{\colhead{UT} & \colhead{$\Delta t$ (days)} &
\colhead{Instrument} & \colhead{Filter/Grating} &  \colhead{Seeing (\arcsec)}
& \colhead{Exptime (s)} & \colhead{Magnitude} }
\startdata
Feb 15.519 & 0.413 & UKIRT/UFTI & K98 & 0.42 & 800&$20.23 \pm 0.11$ \\
Feb 16.497 & 1.391 & UKIRT/UFTI &K98 & 0.48 & 1200&$20.75 \pm 0.22$\\
Feb 17.466 & 2.360 & UKIRT/UFTI & K98 & 1.05&3000 &$21.10 \pm 0.31$\\
Mar 03.565 & 16.457 & GEMINI/NIRI &K &  0.60&2250 &$22.08 \pm 0.29$\\
May  26.381 & 100.274 & GEMINI/NIRI & K & 0.60 & 2250 & $>22.25 (3\sigma)$ \\
\hline
Feb 26.552 & 11.446 & GEMINI/GMOS&r & 1.06  &1800 &24.71 $\pm$ 0.11\\
Mar 8.394 & 21.287 & GEMINI/GMOS&r  & 0.84 & 1800& 24.66 $\pm$ 0.05\\
June 10.287 &  115.180      & GEMINI/GMOS&r & 1.03 &  1800 & 24.68 $\pm$ 0.07 \\
\hline
Mar 11.530 & & GEMINI/GMOS &R400 & - & 7200& -\\
\enddata
\tablecomments{\phantom{}Log of photometry of XRF~050215B obtained at UKIRT and Gemini. 
The times given are the midpoint of the observations, errors quoted are 1$\sigma$.}
\label{phot}
\end{deluxetable*}



\section{Discussion}
Figure \ref{kband} shows the range of magnitudes seen in the K-band afterglow sample
which has been observed to date (Rau \etal\ 2004 and references therein). In addition to 
the sample determined from GRBs discovered before {\it Swift} (e.g. {\it BeppoSAX, HETE-2}) which
are shown in black, the K-band points from afterglows detected by {\it Swift} are also shown 
as dark blue triangles and the highest redshift bursts (GRBs 050814 and 050904) 
are shown in as green squares. The afterglow of XRF~050215B is shown in red and, as can be seen to be the faintest afterglow
observed to date in the K-band (January 2006), and lies toward
the faint end of the X-ray afterglow sample (e.g. Nousek \etal\ 2006; O'Brien \etal\ 2006).

It is possible that the faintness of the optical and X-ray afterglow of XRF~050215B could
be related to the faintness of the prompt emission component. Berger \etal\ (2005) show
that typically {\it Swift} bursts are fainter than those detected by {\it HETE-2} and 
{\it BeppoSAX} at essentially all wavelengths (i.e. both in prompt emission and
it in X-ray/Optical Afterglows). However XRF~050215B was also detected
by {\it HETE-2} and the fluences observed in the 2-30 and 30-400 keV
band were $2.8 \times 10^{-7}$ ergs and $1.7 \times 10^{-7}$ ergs, lying between
the fluences seen for XRF~020903 and 030723, the XRFs which have
the best studied optical afterglows (e.g. Sakamoto \etal\ 2005). 
The extrapolation of the X-ray afterglow of XRF~050215B out to later times,
comparable to those of X-ray observations of previous XRF afterglows, 
show that it would achieve a 10 day flux of approximately $1 \times 10^{-14}$ ergs s$^{-1}$
cm$^{-2}$, somewhat fainter than the afterglows of XRF~011030 and 020427 
at these times (Kouveliotou \etal\ 2004; Levan \etal\ 2005); however it would have been slightly 
brighter than the afterglow of XRF~030723 (Butler \etal\ 2004). The afterglow decay
is, however very shallow - slower than all but four of those studied by O'Brien \etal\ 2006, 
and, furthermore continues to have a shallow decay for $\sim$ 10 days post burst.

A dusty environment does not seem able to explain the faintness of this burst since it
is faint at both X-ray and optical wavelengths. Although dust within the host galaxy
could render the afterglow invisible at optical wavelengths and faint in the IR
it could not simultaneously explain the faintness of the X-ray afterglow. Indeed an
extrapolation of the X-ray afterglow into the optical/IR regime (using the method 
of Rol \etal\ 2005) reveals that for typical fireball parameters the observed K-band
flux and optical limits are consistent with zero extinction. Unfortunately given the lack
of multiband optical observations and the faintness of the X-ray afterglow excess absorption
cannot be searched for either from the shape of the optical SED or by a decrement of
soft X-rays. 

An alternative model is that the burst lies at high redshift. Although it is now 
clear that some XRFs lie at low redshifts (e.g. XRF 020903 at $z=0.25$, Soderberg et al. 2004), it
is plausible that some fraction also originate in the very high redshift universe.
The presence (and indeed
brightness) of the host galaxy of XRF 050215B indicate that
the redshift cannot be very high. 
However moderate luminosity
distance (e.g. z$>2$) may still be a viable explanation of the faintness of the burst, although cannot
explain the observed slow decay rate. Indeed {\it Swift} bursts are apparently 
fainter at all wavelengths than previous samples (Berger \etal\ 2005) and are
also, on average, at significantly higher redshifts (a mean of $z=2.8$; Jakobsson \etal\ 2006). 
It is also interesting to note that the afterglow of XRF
050215B is significantly fainter than that of the high redshift GRBs 050904 ($z=6.29$ -Haislip \etal\ 2006; Price \etal\ 2005; Tagliaferri \etal\ 2005; Kawai \etal\ 2005) 
and GRB 050814 ($z=5.3$ Jakobsson \etal\ 2006). 

The most popular models for the production of XRFs is that they are: (i) GRBs viewed away from
the collimation axis (e.g. Yamazaki \etal\ 2002), (ii) bursts with high baryon contamination
within the jet or (iii) bursts with jets which are intrinsically broader than those in GRBs. 
All of these models can efficiently soften the prompt emission from that seen 
in "classical" GRBs. Although the prompt emission of bursts from these different
mechanisms may be largely indistinguishable, it is possible that their afterglows
will show significant variation in their appearance (e.g. Granot \etal\ 2005). We consider these
models, and how they may apply to the afterglow of XRF~050215B, below. 

\begin{figure}[h]
\plotone{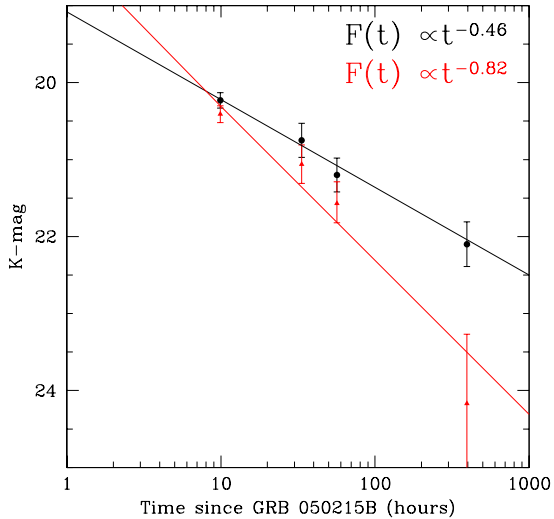}
\caption{The K-band lightcurve of XRF~050215B obtained at UKIRT and Gemini. Also marked
by a heavy line is the best fit power law decay $F(t) \propto t^{-0.47}$, which is very slow for GRB afterglows across this 
time frame. Also shown by triangles and a thin line
is a putative host subtracted light curve (assuming that the host
has K=22.25. The thin line shows a (arbitrarily normalised) powerlaw with the same decay index
as that seen in the X-ray lightcurve of XRF 050215B. As can be seen, within the uncertainty due to host subtraction it is possible that the K band decay has the same index as that seen in X-rays. }
\label{klc}
\end{figure}

\begin{figure}[h]
\plotone{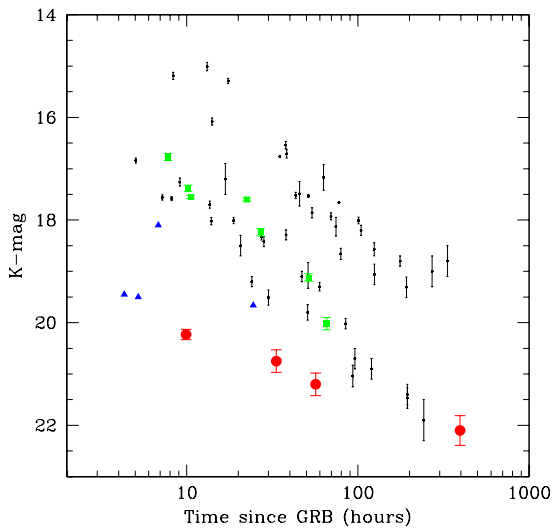}
\caption{A comparison of the afterglow of XRF~050215B with the K-band afterglows 
of other GRBs (modified from that of Rau \etal\ 2004). The red circles are the
afterglow of XRF~050215B, the blue triangles are for {\it Swift} bursts and
the green boxes are for the high-redshift bursts. 
This plot shows that XRF 050215B has the faintest 
K-band afterglow yet seen. The data are taken
from Rau \etal\ (and references therein) except for the more
recent bursts which can be found in Berger \etal\ 2005; 
GRB 050401 Watson \etal\ 2005; GRB 050904 Haislip \etal\ 2006; Tagliaferri \etal\ 2005
and GRB 050814 Jakobsson \etal\ 2006.}
\label{kband}
\end{figure}

In the off-axis model, the highest energy emission (i.e. $\gamma$-rays) is
confined within a narrow cone oriented slightly away from the observer and is not seen. 
The observed isotropic equivalent energy for these bursts ($E_{iso}$)
is thus lower than for
bursts seen on axis and can reproduce the observed relationship between $E_p$ and 
$E_{iso}$ first reported by Amati \etal (2002). 
Granot \etal (2005) consider the afterglows
produced by such a model. The early afterglow of a uniform jet with sharp edges can
be seen to rise at early times as the more energetic material becomes visible; it
then reaches a peak at a time roughly corresponding to the jet break, and from that point,
follows a decay which is indistinguishable from an on-axis GRB (since all
of the ejecta are visible), thus the late time decay slope would be expected
to be rapid $\sim t^{-p}$. The afterglow of XRF~050215B does not show the rapid
rise at early times which may be expected of off-axis models. However the
rapid rise is a feature of jets with sharp edges. More realistic models which
have smooth edges (either intrinsically or due to the interaction of the
jet with the stellar envelope) have a flat (but not necessarily rising) early lightcurve. 
Thus the early observations of XRF~050215B can be explained under this model.
However, the late time slope should be steep ($t^{-p}$) while a shallow slope is
observed to 10 days post burst. Thus indicating that the jet must be moderately wide, 
although other GRBs (e.g. GRB 970508) have been shown to behave as a single
power-law out to large times (see e.g. Bloom, Kulkarni \& Frail 2003). 

Under the dirty fireball model the XRF is produced by a jet which is viewed on-axis, 
but in which the Lorentz factor is reduced due to baryon loading. In this model
the decay of the afterglow is much slower ($f_{\nu} \propto t^{-3(p-1)/4}$ or
$f_{\nu} \propto t^{1/2-3p/4}$ depending of the location of the cooling break (Rhoads 2003)). Thus,
for a typical $p=2$ burst the predicted decay slope would be $t^{-0.75}$, broadly consistent
with the observed temporal slopes. However, dirty fireballs can only create a low
peak energy if the prompt emission is created by external rather than internal shocks. 
Recent observations favour a scenario in which the bursts themselves are caused
by internal shocks (Zhang et al. 2005). Furthermore the shape of the lightcurve, and comparison of 
the XRT and BAT lightcurves place constraints. In the dirty fireball model,
the deceleration occurs when the outflow has collected 1/$\Gamma$ of
rest mass; for lower values of $\Gamma$ (i.e. XRFs) this deceleration time
is significantly longer (scaling as $\Gamma^{-8/3}$). The XRT observations 
cannot be used to constrain the deceleration time since they begin $\sim$ 30 minutes
after the burst. However the shape of the BAT lightcurve implies that 
deceleration must have occurred at early times, and thus that the $\Gamma$ 
cannot have been very low. Observations of future XRFs in X-rays in the seconds
to minutes after the burst will allow stronger constraints to be made. 

Alternatively, under the internal shock model a clean fireball can naturally accommodate
the creation of an XRF (Zhang \& Meszaros 2002), especially if the contrast in
Lorentz factor between shells is small, which leads to
inefficient energy dissipation (e.g. Barraud \etal\ 2004). In this case,
the decay index of $\alpha = 0.82$ can also naturally be accommodated, although only
if it is observed prior to the jet-break. 

A final possibility is that the XRF is caused by the opening angle of the jet itself.
Broad jets in GRBs may be present either alone (e.g. Lamb, Donaghy \& Graziani, 2005)
or as a multicomponent
structure including narrower ($\gamma$-ray emitting) jets (e.g. Berger et al. 2003;
Huang et al. 2004), although such models have recently been shown not to
fit the afterglow of GRB 030329, which they were posited to 
explain (Granot 2005). These broader jets have correspondingly later jet break
times than the narrow emission responsible for the GRB, and would naturally 
explain the lack of a jet break for $>$10 days after the burst.  Indeed, regardless
of the precise mechanism for the generation of this XRF, the lack of jet
break implies that the jet opening angle must be wide, perhaps making
a broader jet the most likely explanation. 

It is also interesting to compare the properties of both the prompt and
X-ray emission of XRF~050215B with those of the population of bursts
detected by {\it Swift}. Although the {\it HETE-2} observations allowed
for this burst to be classified as an XRF, the behaviour as observed by the 
BAT was not significantly different from that seen in many other bursts. 
Indeed the measured photon index of 2.0 is comparable to the mean of 
{\it Swift} bursts in general (e.g. O'Brien \etal\ 2006),
 although it does lie amongst the softer bursts 
detected by BATSE (as do many {\it Swift} GRBs). Thus, it may be that 
{\it Swift}, as for {\it HETE-2} and {\it BeppoSAX}, does detect a
large population of XRFs. However, Swift is unable to accurately constrain
their spectral properties due to its limited bandpass in comparison to
previous missions. Indeed in the case of XRF 050406 the apparent peak
energy $E_p < 15$keV (Romano et al. 2005) and thus although it was 
possible to identify the burst as an XRF its peak energy and detailed
spectral prompt spectral properties could not be accurately measured. Only
XRF 050416 has a measured $E_p$ from {\it Swift}, with
$E_p =15.6$ keV (Sakamoto et al. 2005). However, even in this case
the measurement is relatively weak, since a simple power-law fit 
yields a $\chi^2 / dof < 1$.

The afterglow of the burst
is both fainter than typical (even for {\it Swift} bursts) and also more slowly
declining, with no evidence for a jet break during long observations lasting
for 10 days since the burst. The afterglow is very faint in the 
K-band (fainter than the majority of upper limits on K-band afterglow
brightness) and lies at the faint end of the flux distribution for
long duration GRB detected by {\it Swift}. There are a number
of comparably faint bursts e.g. GRB 050223 - Page \etal\ 2005; 
GRB/XRF 050406 - Romano et al. 2005; 
GRB 050421 - Godet \etal\ 2006, GRB 050911 - Page \etal\ 2006.
It is interesting to note that the proposed explanations for this
faintness vary widely, demonstrating the large range of physical 
processes which affect the brightness of a given afterglow. For
example GRB 050911 was interpreted as being due to a BH-NS merger,
since its lack of an X-ray afterglow was reminiscent of short duration
GRBs (e.g. Gehrels et al. 2005), while its lightcurve could also
be interpreted as being due to several accretion event which are seen
in BH-NS mergers (Davies, Levan \& King 2005). In contrast
GRB 050421 was explained as being due to a "naked" GRB - a burst
occurring in a region of very low density, which thus affected
the brightness of the afterglow (e.g. Taylor et al. 2000).
For XRFs, such explanations are not
normally considered and the most commonly discussed explanations
involve differing geometries of either the jet itself or the observer
orientation. For example the flat slopes of the afterglow lightcurves
are predicted in various off-axis models. However, for "normal" long
duration bursts a generic feature of the afterglow light curve is
a flat phase in the afterglow (e.g. Nousek et al. 2005). This
is normally interpreted as being due to late time energy injection
from the central engine, and is also interpreted as such
in the afterglow of XRF 050406 (Romano et al. 2005), which
has a very flat light curve out to late times $\alpha \sim 0.5$. 
While this flat phase is obviously a generic feature of
X-ray afterglows it is less apparent in the optical and IR. Thus
for XRF 050215B, where the X-ray and optical decays are
similar it is more likely that the relatively slow decay is simply
a result of the jet structure. It is also interesting to note that
other optical afterglows of XRFs have shown similar 
behaviour to XRF 050406 in the optical regime. For example
XRF 020903 showed flaring behaviour and a flat decay
in the R-band (Bersier et al. 2006), behaviour which has
not commonly been seen in GRB afterglows (although the sample
of XRF optical/IR afterglows remains much smaller). 

One of the most important contributions that {\it Swift} has made to the GRB field
comes from the decrease in time between the burst itself and the first pointed
observations. Although in the pre-{\it Swift} era a few bursts were observed
promptly in the optical (e.g. GRB 990123, Akerlof et al. 1999) these were rare,
and prompt, pointed, X-ray observations were never made. It was thus hoped
that the early afterglow emission would provide crucial tests of various
afterglow models. In practice, the behaviour seen is somewhat different
from what was expected (e.g. Tagliaferri et al. 2005; Nousek et al. 2005), and
the presence of frequent X-ray flares (e.g. Burrows et al. 2005) makes the
X-ray afterglows a less "clean" probe than may have been hoped for. Nonetheless,
this early phase can provide powerful diagnostics of various models.
O'Brien et al. (2006) have studied this transition period for a number of 
GRBs and show that generally the extrapolation of the X-ray afterglow to
early times "joins" with the prompt emission and does not exhibit 
a strong discontinuity. This behaviour, when seen in XRFs, 
disfavours a sharp-edged jet if jet geometry is the dominating factor. 
However, for three well studied {\it Swift} X-ray afterglows
(XRF 050215B, XRF 050406 and XRF 050416) there is apparently 
no jet break seen out to late times ($>$ 10 days in each case). Thus, all
of these cases must have wide opening angles. Ultimately,
insight into the XRF phenomena is likely to be made via the subset
of bursts which are detected simultaneously by BAT and another
satellite (e.g. HETE-2, Suzaku-WAM, Konus) and where early
X-ray and optical observations can be paired with well constrained spectral
parameters for the prompt emission.



\section{Conclusions}
We have presented observations of XRF~050215B. This was the first X-ray Flash to be
located by {\it Swift} and was achieved by the use of {\it HETE-2} and {\it Swift} in
synergy, utilising the sensitivity and wide field of view of the {\it Swift-BAT} and
the wide spectral range of {\it HETE-FREGATE}.  The fluence within the 
prompt emission was comparable to that seen in previous XRFs, although the
X-ray lightcurve is fainter than the majority of {\it Swift} bursts, while in the
infrared XRF~050215B is the faintest afterglow ever to have been 
detected (although some bursts have been invisible to comparable limits). Indeed,
XRF~050215B had a flux at 9 hours
which was only just observable with a 4m telescope, and at early times
was invisible to any robotic or UVOT observations.  
As also
suggested by Berger \etal\ (2005), this implies that
locating the afterglows of some GRBs in the {\it Swift} era will require rapid
response observations from larger telescopes since many bursts
will be too faint to be detected by the UVOT or even by larger aperture robotic ground based
telescopes. 

The limited volume of data and lack of redshift available for this burst make it difficult to draw firm
conclusions as to the cause of the soft $\gamma$-ray spectrum; 
however the lack of any observed jet-break to $>$ 10 days post
burst make a broad jet the most likely explanation. The late-time behaviour of
XRF 050215B is qualitatively somewhat similar to the late-time X-ray lightcurve of
the recent GRB 060218,  which, by virtue of its proximity (145 Mpc) had a 
bright X-ray and Optical afterglow. In particular, GRB 060218 also had a long-lasting slow 
decay, as seen in XRF~050215B, although the relative faintness of both afterglow and host galaxy,
coupled with the lack of any supernova emission, imply a much higher redshift in the latter case.  
Furthermore GRB 060218  had an exceptionally long duration in $\gamma$-rays
($t_{90} \sim 2000$s), making the 
status of GRB 060218 relative to the bulk of the XRF population unclear, and pointing 
to the continuing need for further well-observed 
XRFs if we are to understand the diversity in their emission mechanisms and their relationship
to GRBs.



\end{document}